\documentclass[journal]{IEEEtran}
\usepackage{graphicx}
\usepackage{psfrag}
\usepackage[cmex10]{amsmath}
\usepackage{wasysym}
\ifCLASSOPTIONcompsoc
  \usepackage[caption=false,font=normalsize,labelfont=sf,textfont=sf]{subfig}
\else
  \usepackage[caption=false,font=footnotesize]{subfig}
\fi
\begin{document}

\title{Simulation of a  Low-Background Proton Detector for Studying Low-Energy Resonances Relevant in Thermonuclear Reactions}

%
%

\author{David P\'erez-Loureiro~\IEEEmembership{}
        and Christopher Wrede~\IEEEmembership{}
 \thanks{Manuscript received \today. This work was supported by  the U.S. National Science foundation under grant PHY-1102511.}
\thanks{D. ~P\'erez-Loureiro and C.~Wrede are with the National Superconducting Cyclotron Laboratory at Michigan State University, East Lansing, MI 48824 USA (e-mail: perez-loureiro@nscl.msu.edu).}%
\thanks{C.~Wrede is also with the Department of Physics and Astronomy at Michigan State University}%

}

\maketitle
\pagestyle{empty}
\thispagestyle{empty}

\begin{abstract}
A new detector is being developed at the National Superconducting Cyclotron Laboratory (NSCL) to measure  low energy charged-particles from beta-delayed particle emission. These low energy particles are very important for nuclear astrophysics studies. The use of a gaseous system instead of a solid state detector decreases the sensitivity to betas while keeping high efficiency for higher mass charged particles like protons or alphas. This low sensitivity to betas minimizes their contribution to the background down to 150 keV. A detailed simulation tool based on \textsc{Geant4} has been developed for this future detector.
\end{abstract}


\section{Introduction}
%
%
%
%
\IEEEPARstart{C}{lassical} novae and type I x--ray bursters are explosive events that occur in close binary systems containing a compact object, such a white dwarf (novae) or neutron star (x--ray bursters) and an ordinary star.  Due to its gravity, the compact object accretes hydrogen-rich material from its companion which accumulates on its surface. After a sufficient time, the temperature and the pressure at the base of the accumulated material becomes large enough for thermonuclear reactions to occur. These reactions produce an increase in the pressure and temperature, which  leads to runaway and finally an explosion \cite{Iliadis}. Hence, thermonuclear reactions play a very important role not only in the energy generation, but also in the nucleosynthesis of different nuclides in these explosive scenarios.

Due to this dependence, it is important to constrain the reaction rates for better modelling. Most of the reactions of interest are mainly radiative proton captures in which the resonant capture dominates the total rate. In this case, the reaction rate can be evaluated from the information about the energy ($E_{r,j}$) and strength ($\omega\gamma$) of these resonances;  and temperature ($T$) within the xplosion according to \eqref{eq:reaction}:

\begin{equation}
  \label{eq:reaction}
 N_A \langle \sigma v\rangle \propto T^{-3/2} \sum_j(\omega\gamma)_j \exp{\left (\frac{-E_{r,j}}{k_BT}\right )}
\end{equation}

The resonance strengths can be expressed as a function of the spins of the different particles ($J_i$), the total width of the resonance ($\Gamma$) and the partial widths for charged-particle and photon emission ($\Gamma_p,\Gamma_\gamma$) as shown in \eqref{eq:resonance}:

\begin{equation}
  \label{eq:resonance}
  \omega\gamma =\frac{(2J_r+1)}{(2J_{tgt}+1)(2J_p+1)}\frac{\Gamma_p}{\Gamma}\Gamma_\gamma
\end{equation}

\begin{figure}[!t]
\centering
\includegraphics[width=.35\textwidth]{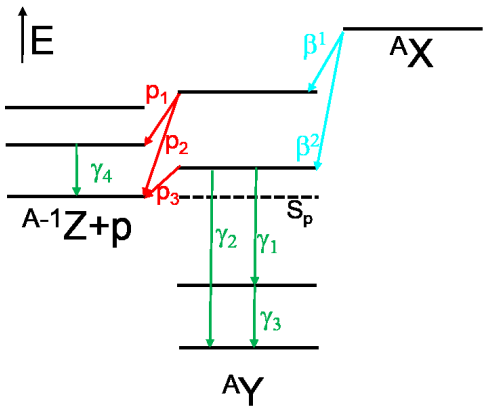}
\caption{Simplified decay scheme of the  beta--delayed proton emission. The precursor ($^\mathrm{A}$X) decays into a excited state of its daughter ($^\mathrm{A}$Y). If this state is above proton separation energy (S$_\mathrm{p}$), it may emit a proton.}
\label{fig_decay}
\end{figure}

The rates of reactions involving stable  nuclei or those with the longest half-lives have been measured directly \cite{Iliadis2002}. Some of the reactions in which unstable nuclides are involved have been investigated in inverse kinematics with rare isotope beams \cite{Ruiz2014}. However, for  short-lived isotopes of certain chemical elements, the beam intensities available are not high enough to make direct studies feasible. The time reversal symmetry makes it possible to investigate these reactions indirectly by studying the same states via beta-delayed proton emission. In this case, a  proton--rich  precursor is produced and left to decay. The Q-value for beta-decay of this precursor is large enough to populate states in the daughter nucleus above the proton emission threshold (Fig.~\ref{fig_decay}). The energy and intensity of the protons emitted can then be measured.

The main challenge in the detection of these protons is that the energy levels which are relevant for nuclear astrophysics studies are the ones closest to the proton emission threshold, meaning that the kinetic energies of the emitted protons are very low. In addition, due to the Coulomb and centrifugal barriers, the proton emission probabilities are very low, competing with electromagnetic de-excitation. When using standard solid state detectors, the energy spectrum of the protons will overlap with the background produced by the positrons emitted during the beta-decay.

\section{Proposed detector}
In order to overcome the difficulty in detecting these low energy protons, a novel detection system has been  developed and tested \cite{Pollacco}. It consists of a gas volume instead of a solid state detector. The radioactive ions are implanted in the gas volume. The gas reduces the sensitivity to the emitted $\beta$--particles pushing the background down in energy. The primary ionization produced by the decay products is amplified by the use of Micro Pattern Gas Amplifier Detectors (MPGADs) like \textsc{Micromegas} \cite{Giomataris}, which assures good energy resolution, efficiency and gain for the detection of protons. A device based on this principle is being designed in the NSCL at Michigan State University for the study of these resonances.The fast rare isotope beams produced by fragmentation in this facility will be implanted in the middle  of the detector to measure its beta--delayed proton emission.

\begin{figure}[!t]
\centering
\includegraphics[width=.45\textwidth]{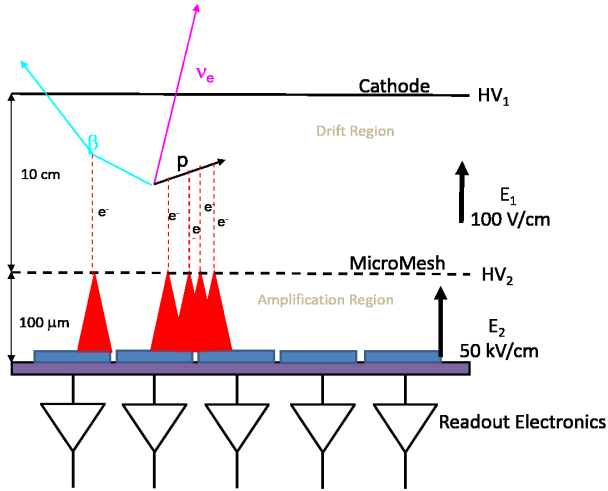}
\caption{Principle of operation of the proton detector based on MPGADs. In this simplified picture, it is assumed the ion decays inside the gas  and no drift of the implanted ions occurs before the decay.}
\label{fig_micromegas}
\end{figure}

\subsection{Principle of operation}
As mentioned before, the detector consists  of a gas volume in which a uniform electric field is applied. When a charged particle traverses the volume, it ionizes the gas, producing electron--ion pairs. Due to this electric field, the  electrons drift towards the amplification region, where a higher gradient will accelerate the electrons creating an avalanche due to secondary ionization. This charge displacement will induce a signal in the readout pads  proportional to the energy of the  particle (Fig.~\ref{fig_micromegas}).

\subsection{Detector Requirements}
The proton detector has to fulfill several requirements: It has to be able to detect the $\beta$--delayed emitted protons or alpha--particles with energies as low as possible with an energy resolution below $\approx$8\% (FWHM), while keeping  the beta background low to minimize its contribution to the energy determination. It also has to be able to be coupled with gamma--ray detectors in order  to distinguish between proton decays in which  gamma--rays are emitted in coincidence from those without electromagentic emission. This will help us to determine whether the proton decay populates  excited states or the ground state in the daughter as depicted in Fig.~\ref{fig_decay}. For certain physics cases, the detector has also to be able to detect multi-particle emission. In order to do that a position sensitive pad plane is foreseen in a future upgrade.

\subsection{Detector Geometry}

\begin{figure}[!t]
\centering
 \psfrag{60}[c][t]{$\diameter$ 60~mm}
 \psfrag{120}{$\diameter$ 120~mm}
\includegraphics[width=.3\textwidth]{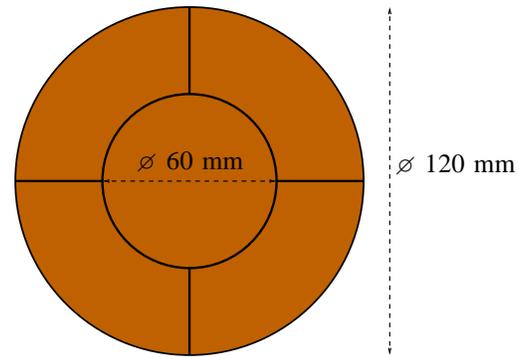}
\caption{Potential geometry of readout pads.}
\label{fig_padGeo}
\end{figure}

The proposed detection system consists of a cylindrical volume with 18~cm diameter and a length of 20~cm. The compact dimensions will allow the system to fit inside the SeGA germanium array  in its barrel configuration \cite{SEGA}, while keeping the symmetry of the whole system. On the downstream end-cap of the tube, the \textsc{micromegas} detector will be placed with one potential pad geometry shown in Fig.~\ref{fig_padGeo}. The readout pad plane is divided into two concentric circles with 6~cm inner diameter and 12~cm  outer diameter. The outer one is divided in four quadrants. This particular configuration will allow anti-coincidences between the outer and inner sectors to be made in roder to veto those particles which did not deposit their full energy in the central pad.


\section{Simulations}

\begin{figure}[!t]
\includegraphics[width=.45\textwidth]{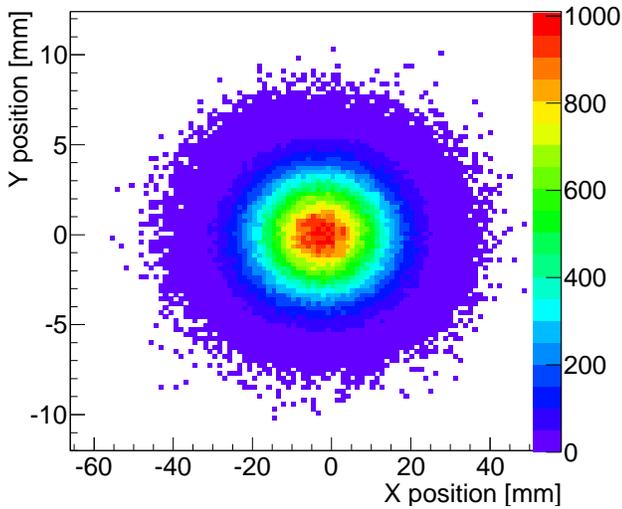}
\caption{Projected position distribution of the implanted  $^{23}$Al ions in the XY plane.}
\label{fig_position}
\end{figure}

The full detector set-up is simulated using a dedicated program based on the {\sc ROOT} data analysis framework \cite{Brun}
and the {\sc Geant4}  toolkit \cite{Geant4}. The program employs {\sc Geant4} standard libraries to describe the interactions of the particles with the atoms or
molecules of the gas and for the determination of the energy deposited at each interaction position along their trajectories. The beta-decay and the proton emission are implemented with a custom physics model for the different ions to be studied, e.g.\ $^{23}$Al or $^{31}$Cl, but it can be adapted to any nucleus of interest. The simulation is  performed in two stages: During the first stage the beam is implanted in the gas volume. In order to have a realistic distribution of the implanted ions, the beam parameters,  i.e.\,, the initial  energy and position distributions of the beam are taken from the results of a LISE++ simulation\cite{LISE}. This simulation accounts for any beam properties, and allows us to make an estimation of any contaminant which will be produced. Figure \ref{fig_position} shows the distribution of the implanted ions in the gas detector projected in the XY plane. We can see  that the distribution is narrower in the Y direction compared to the X direction due to the focusing of the beam. In the second stage, the implanted ions decay nearly at rest in the gas volume. During this stage, the decay products, namely, protons and betas will interact with the gas, leaving part of their kinetic energy. The energy deposition along the whole trajectory is collected during the simulation and stored in a ROOT file on an event--by--event basis. The electron  drift, amplification in the {\sc micromegas}, and the subsequent charge collection on the detection pad plane have also been implemented in a separate program, providing together  a full simulation of the detector, from the event generation up to the charge collection and signal induction.  A description of these processes, how they are combined with the energy deposition given by  {\sc Geant4} and position information to simulate the detector response can be found in \cite{Pancin}. It essentially calculates the number of electrons produced in each ionization step along the trajectory of the particle through the gas and randomizes its final position in the anode plane. During the amplification stage the noise and the fluctuations in the avalanche are also taken into account to get a realistic detector response. The charge collected in each pad is the sum of all the electrons which reach the pad after the amplification.

\begin{figure}[!t]
\centering
\includegraphics[width=2.75in,height=2.75in]{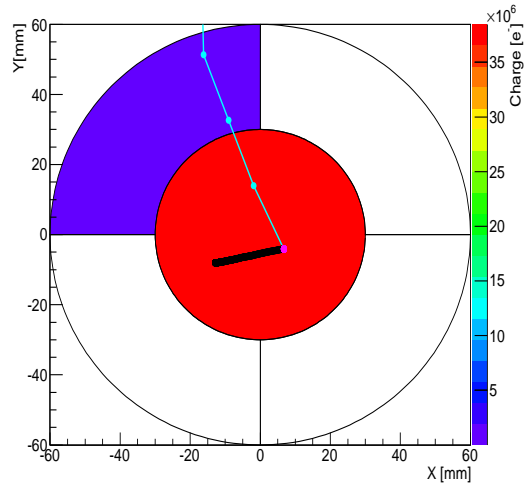}
\caption{Sample of a beta-decay event and the subsequent proton emission event projected on the perpendicular plane to the beam axis. The blue and the black dots correspond to the projection of the trajectories of the betas and the protons respectively. The color scale in the histogram shows the charge deposited in each pad in elementary charge units.}
\label{fig_pads}
\end{figure}

Fig. \ref{fig_pads} shows a sample of a decay event projected in the plane perpendicular to the beam axis. We can see the  simulated traces of the proton and the beta-particle. We observe that the proton (black circles) is completely stopped in the central pad of the detector depositing its full energy in it. However, the beta only deposits a small fraction of its energy in the different sectors of the pad plane and escapes the  active region of the detector.

\subsection{Results}
\begin{figure}[!t]
\includegraphics[width=.45\textwidth]{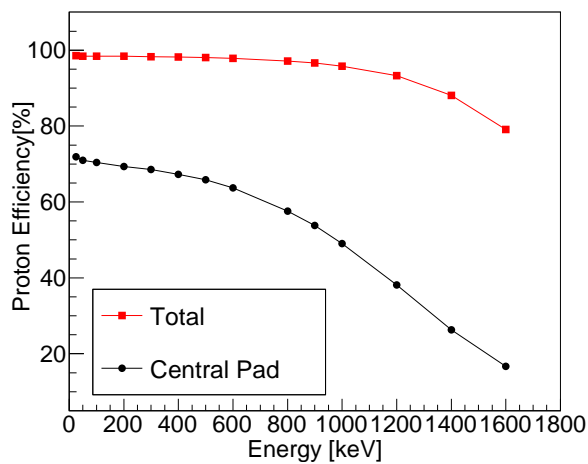}
\caption{Full energy proton efficiency versus proton kinetic energy for the central pad (circles) and the full detector (squares) for a mixture of Ar/CH$_4$ (95\%-5\%) at 800 torr.}
\label{fig_eff}
\end{figure}

In order to investigate the performance of the proposed detector we performed different simulations.
One of the main requirements is the high efficiency. Fig.~\ref{fig_eff} shows the evolution of the efficiency with the kinetic energy of the protons emitted during a decay. The detection gas is a mixture of Ar/CH$_4$ (95\%-5\%) at 800 torr. Again, it is assumed that ions are stopped in the gas volume and they do not drift towards the cathode before decaying. As we can see, the efficiency is above 90\% for energies below 1.4~MeV for the whole detection system (squares). Above that energy, the range of the protons in the gas becomes larger and some of them will escape the active volume. In the case of only the central pad (circles), the efficiency is above 60\% below 700 keV. This is due to the fact that the initial distribution of the implanted ions is wider than the pad size as  shown in  Fig. \ref{fig_position} . Of course, this can be improved up to close to 100\% by collimating our beam in the X direction but losing some of the initial beam intensity. For higher energies, it rapidly drops below 20\%. This is explained by the fact that the range is larger than the central pad size and most of the protons will traverse more than one pad, sharing their energy. This efficiency can be improved by increasing the central pad size or the gas pressure to decrease the range of the protons, however, this will increase  the sensitivity to $\beta$--particles which will produce a higher contribution to the  background.

\begin{figure}[!t]
\centering
\includegraphics[width=.45\textwidth]{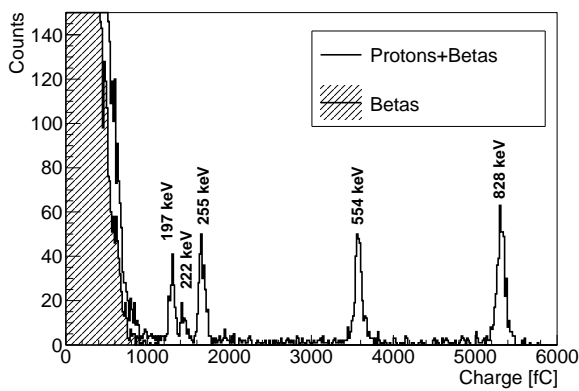}
\caption{Simulated spectrum of the central pad of the proton detector for the $\beta$-decay of $^{23}$Al. The shaded histogram corresponds to the contribution of the betas while the black line shows the sum of both, protons and the betas. In this spectrum anticoincidences between inner and outer pads have not been applied.}
\label{fig_spectrum}
\end{figure}

As far as the energy resolution is concerned, Figure ~\ref{fig_spectrum} shows the spectrum obtained from a simulation of the beta decay of $^{23}$Al. The decay of this nucleus is a good candidate for investigating the performance of this detector due to the known feeding of levels only 200~keV above the proton threshold which emit protons \cite{Saastamoinen}. The black line shows the sum of all contributions, that is protons and betas. As it is shown, we can distinguish protons with energies as low as 197~keV and rather low intensities with a resolution better than 7\% FWHM below 400 keV. The shaded spectrum corresponds to the contribution due to betas. We can see that the beta background is almost negligible above 950~fC, which correspond to an energy around 150~keV, as required for astrophysical applications.

\section{Conclusion}
A {\sc Geant4} based program  has been developed for the simulation of a new particle detector to be used at NSCL for nuclear astrophysics studies relevant for nova nucleosynthesis.
This detector  based on MPGADs, compared to solid-state detection systems, will allow us to measure beta-delayed protons and alphas with energies as low as 200~keV even at very low intensities with a very low background due to betas.
The simulation developed describes the performance of MPGADs, showing that  the energy deposition due to betas can be pushed below 150~keV. Indeed, this has already been demonstrated using an similar detector at another facility, the ASTROBOX prototype.


\section*{Acknowledgment}
The authors gratefully acknowledge E. Pollacco and the AstroBox collaboration  for their valuable help and fruitful discussions.


\bibliographystyle{IEEEtran}
\bibliography{IEEEabrv,bibliography}
%




\end{document}